\documentclass[preprint,12pt,authoryear]{elsarticle}

\newcommand{\apj}{ApJ}

\newcommand{\aap}{A\&A}

\newcommand{\simgt}{\lower 2pt \hbox{$\, \buildrel {\scriptstyle >}\over {\scriptstyle\sim}\,$}}
\newcommand{\simlt}{\lower 2pt \hbox{$\, \buildrel {\scriptstyle <}\over {\scriptstyle\sim}\,$}}

\newcommand{\xmm}{\emph{XMM-Newton}}
\newcommand{\suz}{\emph{Suzaku}}

\usepackage{graphicx}
\usepackage{epsfig}

\usepackage{amssymb}

\usepackage[colorlinks=true, linkcolor=blue, citecolor=red, urlcolor=blue]{hyperref}
\usepackage{natbib}
\bibpunct{(}{)}{;}{a}{}{,}

\journal{New Astronomy Reviews}

\begin{document}

\begin{frontmatter}



\title{High Velocity Outflows in Narrow Absorption Line Quasars}


\author{G. Chartas, J. Charlton, M. Eracleous, M. Giustini, P. Rodriguez Hidalgo}
\address{Department of Astronomy \& Astrophysics, Pennsylvania State University,
University Park, PA 16802, chartas@astro.psu.edu}
\author{R. Ganguly}
\address{Department of Computer Science, Engineering, \& Physics,
University of Michigan-Flint, Flint, MI 48502}
\author{F. Hamann}
\address{Department of Astronomy, University of Florida, 211 Bryant Space Science Center, Gainesville, FL 32611-2055}
\author{T. Misawa}
\address{Cosmic Radiation Laboratory, RIKEN, 2-1 Hirosawa, Wako, Saitama 351-0198 Japan}
\author{D. Tytler}
\address{Center for Astrophysics and Space Sciences, University of California San Diego, La Jolla, CA 92093-0424, USA}

\address{}

\begin{abstract}

The current paradigm for the AGN phenomenon is a
central engine that consists of an inflow of material accreting
in the form of a disk onto a supermassive black hole.
Observations in the UV and optical find high velocity
ionized material outflowing from the black hole.
We present results from \suz\ and  \xmm\ observations ofÊ
a sample of intrinsic NAL quasars with high velocity outflows.
Our derived values of the intrinsic column densities of the X-ray absorbers
are consistent with an outflow scenario in which NAL quasars 
are viewed at smaller inclination angles than BAL quasars. 
We find that the distributions of $\alpha_{\rm OX}$ and 
$\Delta\alpha_ {\rm OX}$ of the NAL quasars of 
our sample differ significantly from those of BAL quasars and SDSS radio-quiet quasars. The NAL quasars are not 
significantly absorbed in the X-ray band and the positive values of  
$\Delta\alpha_ {\rm OX}$ suggest 
absorption in the UV band.
The positive values of $\Delta\alpha_ {\rm OX}$ of the intrinsic NAL quasars can be 
explained in a geometric scenario where our lines of sight towards the compact 
X-ray hot coronae of NAL quasars do not traverse the absorbing wind 
whereas lines of sight towards their UV emitting accretion disks do intercept 
the outflowing absorbers. 

\end{abstract}


\end{frontmatter}



\section{Introduction}
\label{intro}

Optical and UV absorption lines in quasars are commonly classified by their widths
into ``broad'' (BALs; FWHM $ >$ 2,000~km~s$^{-1}$),
``narrow'' (NALs; FWHM $ $\simlt$ $500~km~s$^{-1}$), 
and mini-BALs with absorption line widths ranging between those
of BALs and NALs. 
These class definitions are considered somewhat arbitrary.
The definition of NALs for example was chosen such that the 
the C~IV doublet can be resolved.
Several models of quasar structure indicate that the widths of intrinsic 
absorption lines may depend on the angle between our line of sight and that of 
the outflowing absorbing stream, and on the velocity gradient in the outflowing stream.

UV spectroscopic observations have revealed highly blueshifted narrow 
and broad intrinsic absorption lines in quasars implying outflow velocities of 
up to $\sim$ 60,000~km s$^{-1}$ \citep[e.g.,][]{jannuzi96,hamann97,narayanan04}.
Intrinsic NALs are common in Type I AGN, occurring in $\sim$50\% of 
of optically selected quasars \citep{misawa07}.
They may be present in all AGN but only detected in those cases where our 
line of sight intersects the outflowing absorbing stream.
Most of our current understanding of the physical and kinematic
structure of NALs and BALs stems from studies of the velocity profiles 
of absorption lines that appear blueward of resonance UV emission lines.
Little is presently known about the absorbing and kinematic properties
of these absorption line systems in the X-ray band. 

In section \S 2 we describe the 
NAL Quasar sample selection, 
and in \S 3 the observations, data analysis and results 
of the sample.
Finally in section \S 4 we conclude by summarizing the results of our observations
of NAL quasars.

Throughout this paper we adopt a $\Lambda$-dominated cosmology with 
$H_{0}$ = 70~km~s$^{-1}$~Mpc$^{-1}$, 
$\Omega_{\rm \Lambda}$ = 0.7, and  $\Omega_{\rm M}$ = 0.3.

\section{NAL Quasar Sample Selection}

Our initial NAL quasar sample consisted of $z$=2--4 quasars observed with 
either the Keck telescope at high spectral resolution for the original purpose of
studying intergalactic deuterium lines or quasars obtained from the VLT/UVES archive. The VLT/UVES observations
were taken at high spectral resolution (R $\sim$ 40,000) for various scientific purposes.
From this sample we identified intrinsic NALs by the partial coverage signature of their C~IV, N V and S IV
absorption doublets \citep{misawa07}.
The X-ray sample used in this paper consists of 12 $z=2.2$--3.1 quasars 
from the initial optical sample that contained intrinsic NALs. 
The X-ray NAL sample was selected from the initial NAL sample by choosing
sources bright enough in the UV resulting in
at least 100 counts in the {\sl Suzaku} and/or {\sl XMM-Newton} observations (see Table 1).
We have also included in our analysis the 4 NAL quasars from the exploratory survey of  
\cite{misawa08}.  
The X-ray sample does not contain any BAL quasars.
Thus our sample consists of secure intrinsic NALs 
at high ejection velocities, (63,000~km/s $ \simgt $ $v_{\rm ej}$ $ \simgt $ 6,400~km/s).
More importantly, the sample is unbiased, selected
without regard to quasar properties. Thus it is representative of intrinsic
NALs and lends itself to drawing general conclusions about
the properties of the absorbers. We note a possible luminosity bias in 
the X-ray selected sample since we selected the brightest objects from the initial NAL quasar sample 
to obtain moderate S/N ratio X-ray spectra.

\section{X-ray Observations of NAL Quasar Sample}

A log of the observations that includes 
object, observation dates, observatory, observed count rates, 
total exposure times, 
and observational identification numbers is presented in Table 1. 

\begin{table}[t]
\scriptsize
\begin{center}
\begin{tabular}{llcccc}
 & & & &&\\ \hline\hline
                  &&                                        & & Effective  &      \\
Object&Observation & Observatory    &  Observation  &    Exposure Time${}^{a}$  & $N_{sc}$${}^{b}$\\
&Date           &                      &  ID                  & (ks)     &     \\
\hline
\hline
Q0109$-$3518    & 2008 May 20            & {\it Suzaku}      & 703037010  &  21.11 &  332 $\pm$ 18  \\
Q0122$-$380  & 2008 May 29               & {\it Suzaku}       & 703035010   & 19.14   & 217 $\pm$ 15  \\
Q0329$-$255      & 2008 June 16             & {\it Suzaku}   &  703038010  & 21.0&  784 $\pm$ 28 \\
Q0450$-$1310  & 2008 March 10             & {\it Suzaku}          & 702062010   & 7.5 & 106 $\pm$ 10 \\ 
Q0450$-$1310  &2007 August 10    & {\it XMM-Newton}  & 0503350301  & 5.88  & 242 $\pm$ 16   \\
Q0551$-$3637   & 2008 May 14              & {\it Suzaku}                 &703036020   & 8.5  & 127 $\pm$ 11  \\
Q0940$-$1050      & 2008 May 5               & {\it Suzaku}         & 703040010   & 18.8   &  289 $\pm$ 52  \\
Q1009+2956 &2007 October 31   & {\it XMM-Newton}  & 0503350201   & 5.69  &  487 $\pm$ 22 \\
Q1017+1055 &2007 November 27    & {\it Suzaku}          &702064010  &  8.13 & 131 $\pm$ 12   \\
Q1158$-$1843     & 2008 June 19               & {\it Suzaku}        & 703039010  & 7.5  & 307 $\pm$ 41  \\
Q1334$-$0033 &2007 July 14              & {\it Suzaku}          & 702067010 & 12.11 & 231 $\pm$ 15  \\
Q1548+0917 &  2008 Feb 02               & {\it Suzaku}       & 702068010  &  30.48 & 612 $\pm$ 25   \\
Q1946+7658 &2007 July 13             & {\it Suzaku}          & 702060010  & 11.66 & 398 $\pm$ 20  \\
Q1946+7658 &2007 July 11        &  {\it XMM-Newton}  &  0503350101  &2.96 &  200 $\pm$ 14   \\
\hline \hline
\end{tabular}
\end{center}
\caption{ Log of Observations of our \suz\ and \xmm\ NAL Quasar Sample.
${}^{a}$Effective exposure time is the time remaining after the application of good time-interval (GTI)
tables to remove portions of the observation that were severely contaminated by background.
${}^{b}$ Background-subtracted source counts including events with energies within the 0.2--10~keV band.
The source counts and effective exposure times for the \suz\ and \xmm\ observations refer to those obtained with the
combined XIS units and EPIC PN instrument, respectively. }
\end{table}

In Table 2 we list the redshift, luminosity distance, Galactic column density,
flux density at 2500\AA, outflow velocity and the total rest-frame equivalent width ($W_{\rm rest}$) of the C IV absorption line.

\begin{table}[t]
\scriptsize
\begin{center}
\begin{tabular}{llcccccc}
 & & & & &&&\\ \hline\hline
                  &&                                &       & &  &&    \\
Quasar    & $z$     &  $D_{\rm L}$${}^{a}$   &    $N^{\rm Gal}_{\rm H}$ & $f_{\rm 2~keV}$${}^{b}$& $f_{\rm 2500 \AA}$ ${}^{c}$& $v_{\rm shift}$${}^{d}$ & Rest EW  \\
              &           & Gpc             &   10$^{20}$~cm$^{-2}$   &erg~s$^{-1}$~cm$^{-2}$~keV$^{-1}$ &   mJy   & km~s$^{-1}$  &  \AA        \\
\hline
\hline
Q0109$-$3518    & 2.405 &  20.0&  1.93  &1.2 $\times$ 10$^{-13}$ &0.261   & $-$62,990 &0.064   \\
Q0122$-$380      & 2.2     &  18.0& 1.77   &4.3 $\times$ 10$^{-14}$& 0.171  & $-$40,613 & 0.025  \\
Q0329$-$385      &  2.423  & 20.2 &  1.60  &1.1 $\times$ 10$^{-13}$& 0.154 & $-58,879$ &  \\
Q0450$-$1310    & 2.300   & 19.0 & 6.44    &3.2 $\times$ 10$^{-14}$& 0.185   &$-$6,369  & 0.17  \\
Q0551$-$3637    & 2.318   & 19.1& 3.15   &5.2 $\times$ 10$^{-13}$&0.165   & $-$51,179 &0.3   \\
Q0940$-$1050    & 3.080  & 27.1&  4.17  &8.5 $\times$ 10$^{-14}$& 0.203  & $-$18,576 &0.14   \\
Q1009+2956   &  2.644  & 22.5& 2.40   &1.5 $\times$ 10$^{-13}$& 0.277  & $-$33,879 & 0.1 \\
Q1017+1055   & 3.156   & 27.9 & 3.66    &1.4 $\times$ 10$^{-13}$&0.083   &$-$47,660 &4.92  \\
Q1158$-$1843    & 2.448   & 20.5 & 3.80   &3.0 $\times$ 10$^{-13}$& 0.167  & $-$61,545 & 0.086   \\
Q1334$-$0033   & 2.801    & 24.1& 2.02   &3.9 $\times$ 10$^{-13}$& 0.122  &$-$51,040  &0.36   \\
Q1548+0917  & 2.749     & 23.6&  3.43  &6.2 $\times$ 10$^{-14}$&  0.057 & $-$36,296  &0.32   \\
Q1946+7658  & 3.051     & 26.8 &  7.58  &1.4 $\times$ 10$^{-13}$& 0.197   &$-$11,940  & 0.1  \\
\hline \hline
\end{tabular}
\end{center}
\caption{The NAL Quasar Sample.
${}^{a}$ The luminosity distance, computed based on the cosmological parameters given at the end of $\S$ 1 of the text.
${}^{b}$ Flux density at 2~keV in the quasar rest-frame. Flux densities are derived from the {\sl Suzaku} observations listed in Table 1
with the exception of Q1009+2956 where we use the {\sl XMM-Newton} observation of this object.
${}^{c}$ Flux density per unit frequency at 2500 {\AA} in the quasar rest-frame, derived from the V- or R-band flux
(assuming $f_{\nu}$ $\propto$ ${\nu}^{-0.44}$ ) and corrected for Galactic extinction but not intrinsic extinction.
${}^{d}$ Velocity offset of a NAL relative to the redshift of the quasar. A negative value denotes a blueshift.}
\end{table}

Eleven NAL quasars were observed with {\sl Suzaku} and  three with {\sl XMM-Newton}. The {\sl Suzaku} data were collected with the two 
front-illuminated (FI) CCDs XIS 0, 3 and one back illuminated
CCD XIS 1.  The {\sl XMM-Newton} data were collected with the EPIC pn and MOS instruments.
We used standard {\sl Suzaku} and {\sl XMM-Newton} processing pipeline
to reduce the X-ray data.

The X-ray spectra were fitted 
simultaneously with two models employing \verb+XSPEC+ version 12 \citep{arnaud96}. 
Specifically, the spectral models used in our analysis are the following: 
(a) A simple power-law, and  (b) a simple power-law with intrinsic absorption.
All models contain absorption due to our Galaxy.

In Figure 1 (left panels) we show the XIS spectra of three quasars of our sample (first three in Table 1)
fit with a model consisting of a power law modified by intrinsic absorption (model b)
together with $\chi^{2}$ residuals of the fits. The right panels of Figure 1 show 
68\%, 90\% and 99\% $\chi^{2}$ confidence contours of $N_{\rm H}$ versus the photon index, $\Gamma$.
We find that for all NAL quasars of our sample the $\chi^{2}$ confidence contours imply no significant intrinsic
absorption with the possible exceptions for quasars Q0109-3518 and Q1158-1843 where there is a marginal detection of an intrinsic column density of 
$N_{\rm H}$ $\sim$ a few times 10$^{22}$~cm$^{-2}$.

\begin{figure}[h!]
\centering
\includegraphics[width=1.0\textwidth]{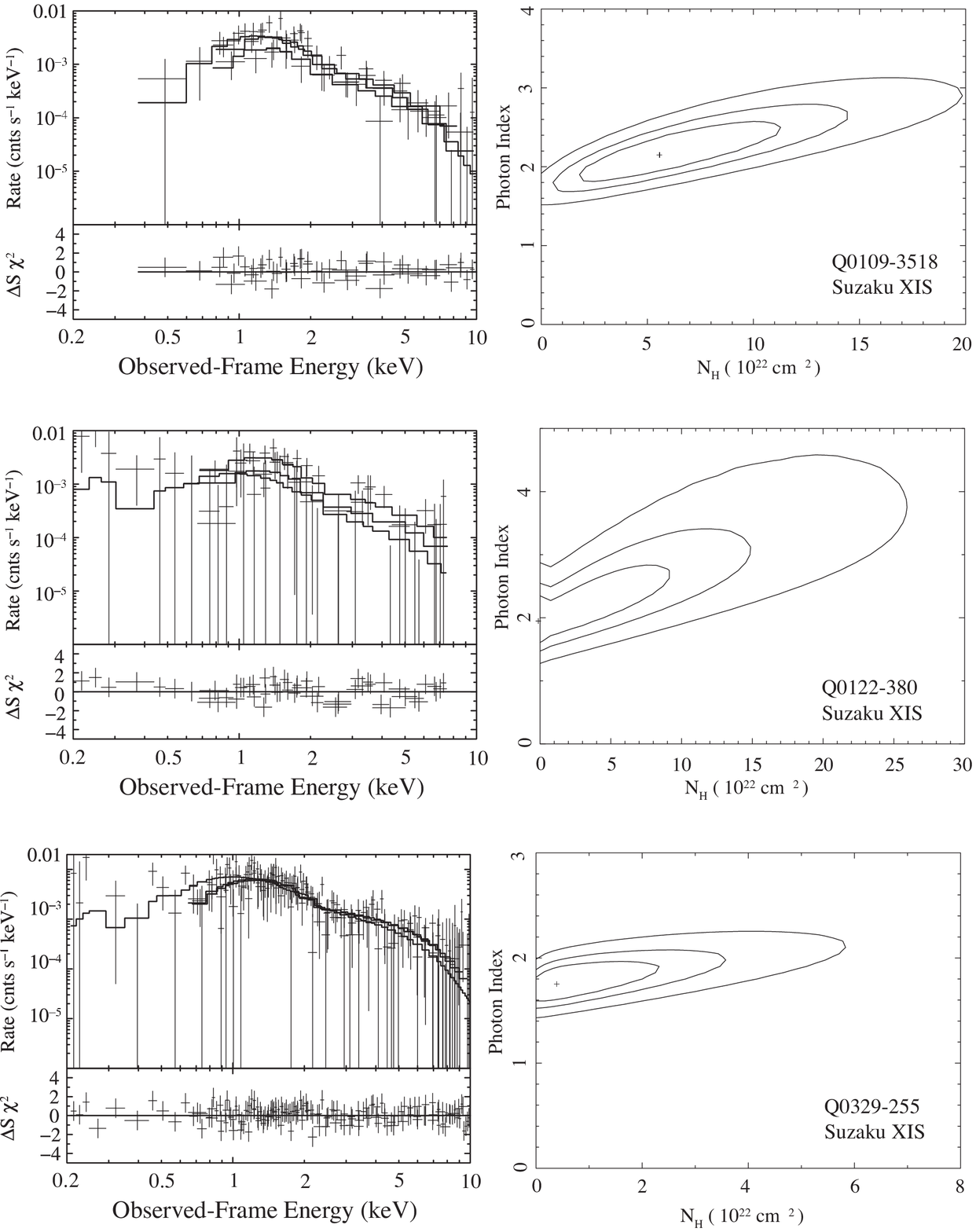}
\caption{In the left panels we present the XIS spectra of three quasars of our sample fit with a 
model consisting of a power-law modified by intrinsic absorption (model b)
together with $\chi^{2}$ residuals of the fits. The right panels show the 
68\%, 90\% and 99\% $\chi^{2}$ confidence contours of the intrinsic column density, 
$N_{\rm H}$, versus the photon index, $\Gamma$.}
\label{spectra}
\end{figure}

In Figure 2 we show the distribution of values of $\alpha_{\rm ox}$ of NAL quasars
without correction for intrinsic UV absorption,
where, $\alpha_{ox}$ is the
optical-to-X-ray slope $\alpha_{ox}$ =$ 0.384\log(f_{2keV}/f_{\rm 2500\;\AA})$ 
\citep{tananbaum79}.
Twelve quasars in this sample are from this study and four are from the \cite{misawa08} study. 
The dashed line shows the distribution of type 1 radio-quiet quasars with
$l_{2500}$ values similar to those of our sample \citep[see Table 5 of~][]{steffen06}.
We notice a significant shift of the  $\alpha_{\rm ox}$  distribution  of NAL quasars with 
respect to the optical-selected type 1 SDSS quasars. 
One possible interpretation of this shift is that on average 
NAL quasars may be more UV absorbed than type 1 SDSS quasars.

\begin{figure}[h!]
\centering
\includegraphics[width=1.0\textwidth]{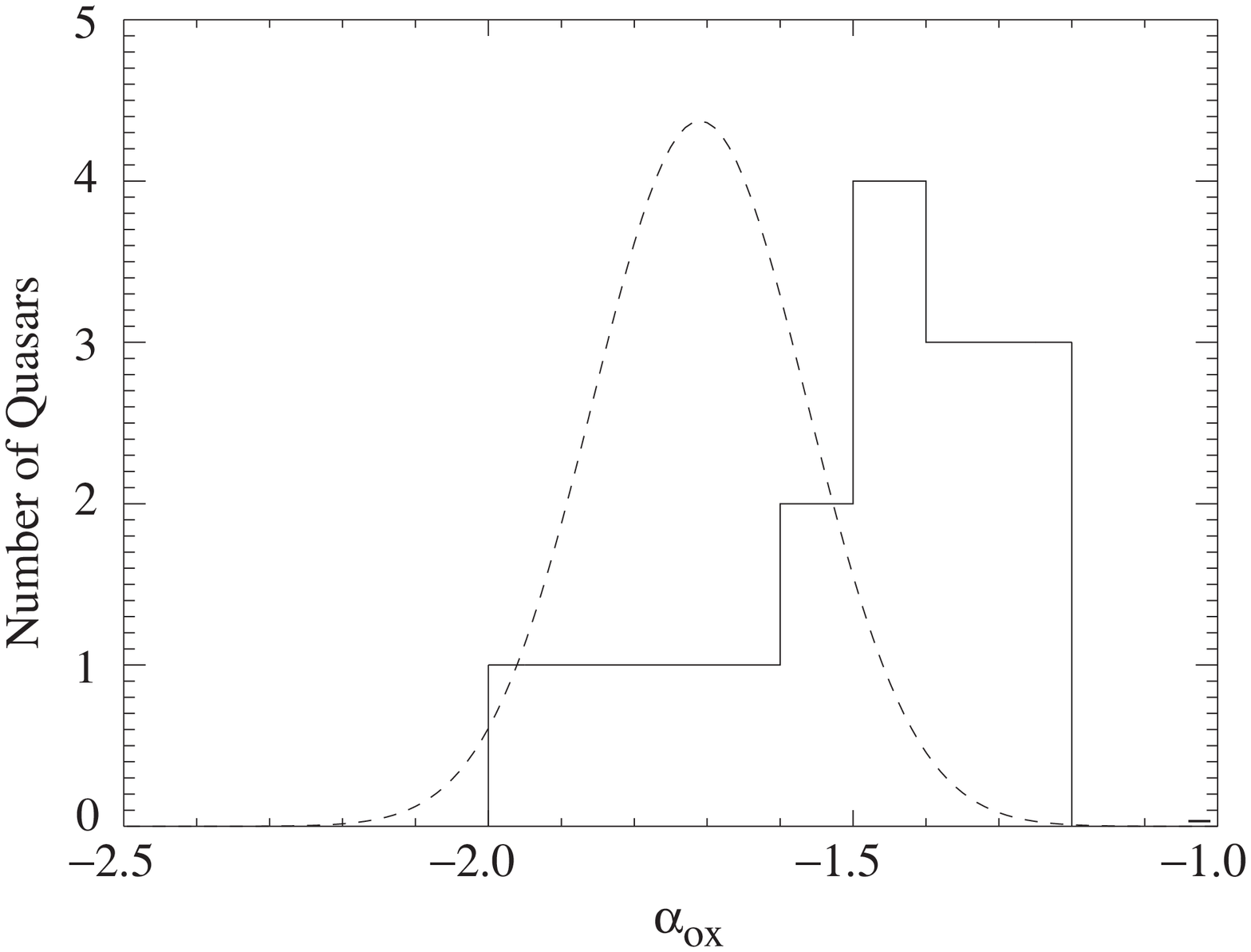}
\caption{The distribution of values of $\alpha_{\rm ox}$ of NAL quasars 
without correction for intrinsic UV absorption. Twelve quasars in this sample are 
from this study and four are from the \cite{misawa08} study. 
The dashed line shows the distribution of type 1 SDSS radio-quiet quasars with
$l_{2500}$ values similar to those of our sample \citep{steffen06}.
A significant shift of the  $\alpha_{\rm ox}$  distribution of NAL quasars with respect to 
the type 1 SDSS radio-quiet quasars may indicate that on average NAL quasars are more UV absorbed than the SDSS quasars.}
\label{aox}
\end{figure}

Since $\alpha_{ox}$ is known to correlate with UV luminosity \citep[e.g.,][]{at86}
we also calculate the parameter $\Delta\alpha_{ox}$ = $\alpha_{ox}$ - $\alpha_{ox}(\ell_{\rm 2500\;\AA})$, where
$\alpha_{ox}(\ell_{\rm 2500\;\AA})$ is the expected $\alpha_{ox}$ for the
monochromatic luminosity at 2500{\rm \AA} \citep[Eq. 6 of~][]{strateva05}.
$\Delta\alpha_{ox}$ is a proxy of X-ray
weakness corrected for the dependence of $\alpha_{ox}$ on UV
luminosity. 

In Figure 3 we show the distribution of $\Delta\alpha_{ox}$
for our current NAL sample, the Steffen et al. 2006 SDSS quasar sample, 
the \cite{giustini08} BAL quasar sample
and the \cite{gallagher06} LBQS BAL quasar sample.
Interestingly our preliminary results hint towards 
positive values of $\Delta\alpha_{ox}$ for NAL quasars.

The positive values of $\Delta\alpha_ {\rm OX}$ of the intrinsic NAL quasars
can be explained in a geometric scenario where NAL quasars are viewed at low 
inclination angles. A possible geometric configuration that can explain this is shown in Figure 4.
In this scenario, lines of sight towards the 
compact X-ray hot coronae of NAL quasars do not traverse the absorbing wind 
whereas lines of sight towards their UV emitting accretion disks do intercept the outflowing absorbers.
Objects that are viewed along lines-of-sight that transverse a substantial portion of the outflowing wind
will appear to be Compton thick.

In Figure 5 we show the total rest-frame CIV equivalent width, $W_{\rm rest}$ (top panel), 
and the maximum NAL velocities of intrinsic CIV NALs of quasars (bottom panel) in our sample (filled circles) and the 
\cite{misawa08} sample (open circles) plotted against $\alpha_{ox}$ (evaluated without corrections 
for intrinsic absorption).  $W_{\rm rest}$ vs. $\alpha_{ox}$  in intrinsic NAL quasars appears to follow 
the trend of $W_{\rm rest}$ vs. $\alpha_{ox}$ found by \cite{brandt00} in low-redshift quasars.
It is interesting to point out that the only object from our sample
that does not appear to follow this trend is Q1017+1055 that contains both a NAL and a mini-BAL.
We finally note that the maximum outflow velocities of the UV absorbers of intrinsic NAL quasars 
do not appear to be correlated with their X-ray weakness in contrast to what has been reported for
BAL quasars \citep[i.e.,][]{gallagher06}.

\begin{figure}[h!]
\centering
\includegraphics[width=1.0\textwidth]{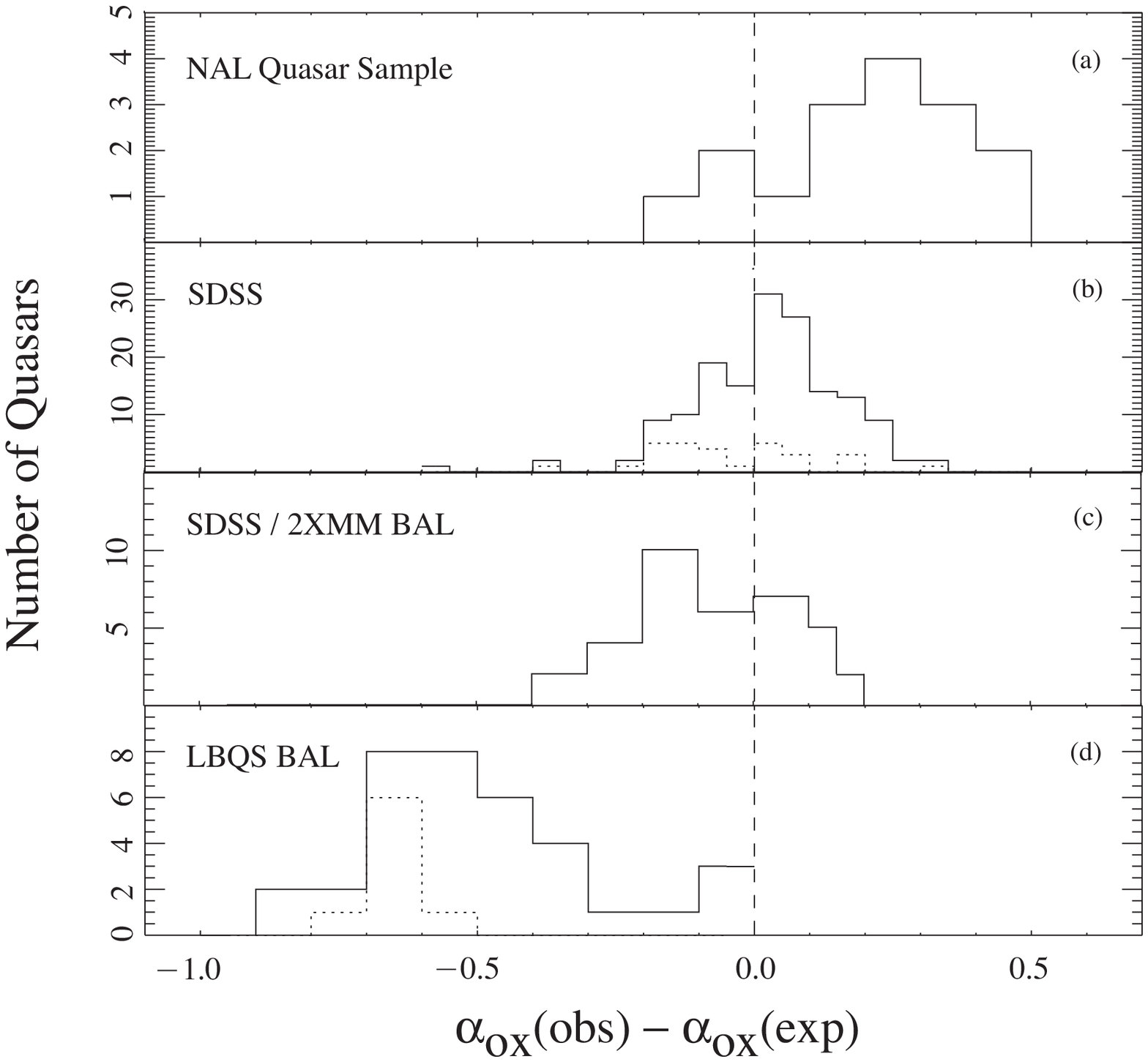}
\caption{Distribution of $\Delta\alpha_{\rm ox}$, the difference between the observed value of $\alpha_{\rm ox}$
and the value predicted for that monochromatic UV luminosity by the correlation of \cite{steffen06}.
Negative values of $\Delta\alpha_{\rm ox}$ indicate a steeper slope than expected.
Histograms drawn as dotted lines indicate upper limits.
{\it (a)}: The distribution of $\Delta\alpha_{\rm ox}$ among quasars in our sample and the \cite{misawa08}    
(without corrections for intrinsic absorption).
{\it (b)}: The distribution of $\Delta\alpha_{\rm ox}$ among SDSS quasars from \cite{steffen06}. 
{\it (c)}: The distribution of $\Delta\alpha_{\rm ox}$ among the \cite{giustini08} BAL quasar sample. 
{\it (d)}: The distribution
of  $\Delta\alpha_{\rm ox}$ among BAL quasars from the LBQS \citep[from][]{gallagher06}.}
\label{daox}
\end{figure}

\begin{figure}[h!]
\centering
\includegraphics[width=1.0\textwidth]{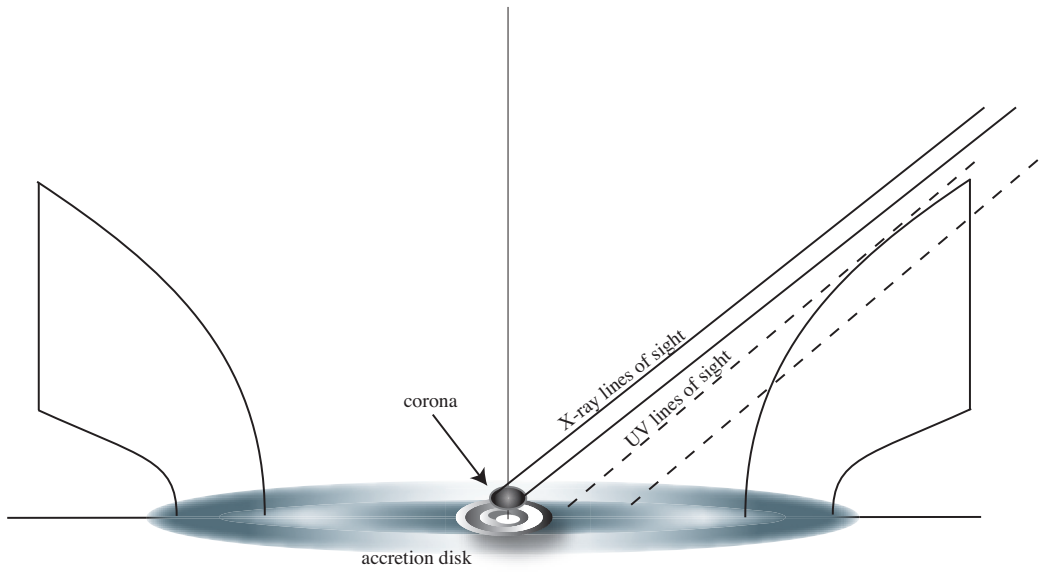}
\caption{Schematic diagram of a proposed geometry for the 
accretion disk and associated outflow in quasars. 
For large inclination angles X-ray emission from the near side of the 
accretion disk and the central continuum source is blocked by the 
Compton thick absorbing wind. 
X-ray lines of sights originating from the corona are indicated
with solid lines and UV lines of sight originating from the accretion disk are indicated
with dashed lines.
Scattered and fluorescent emission
from the far side of the accretion disk and outflow may reach the observer.
Light rays that originate near the black hole will be slightly bent due to GR effects.}
\label{cartoon}
\end{figure}

\begin{figure}[h!]
\centering
\includegraphics[width=1.0\textwidth]{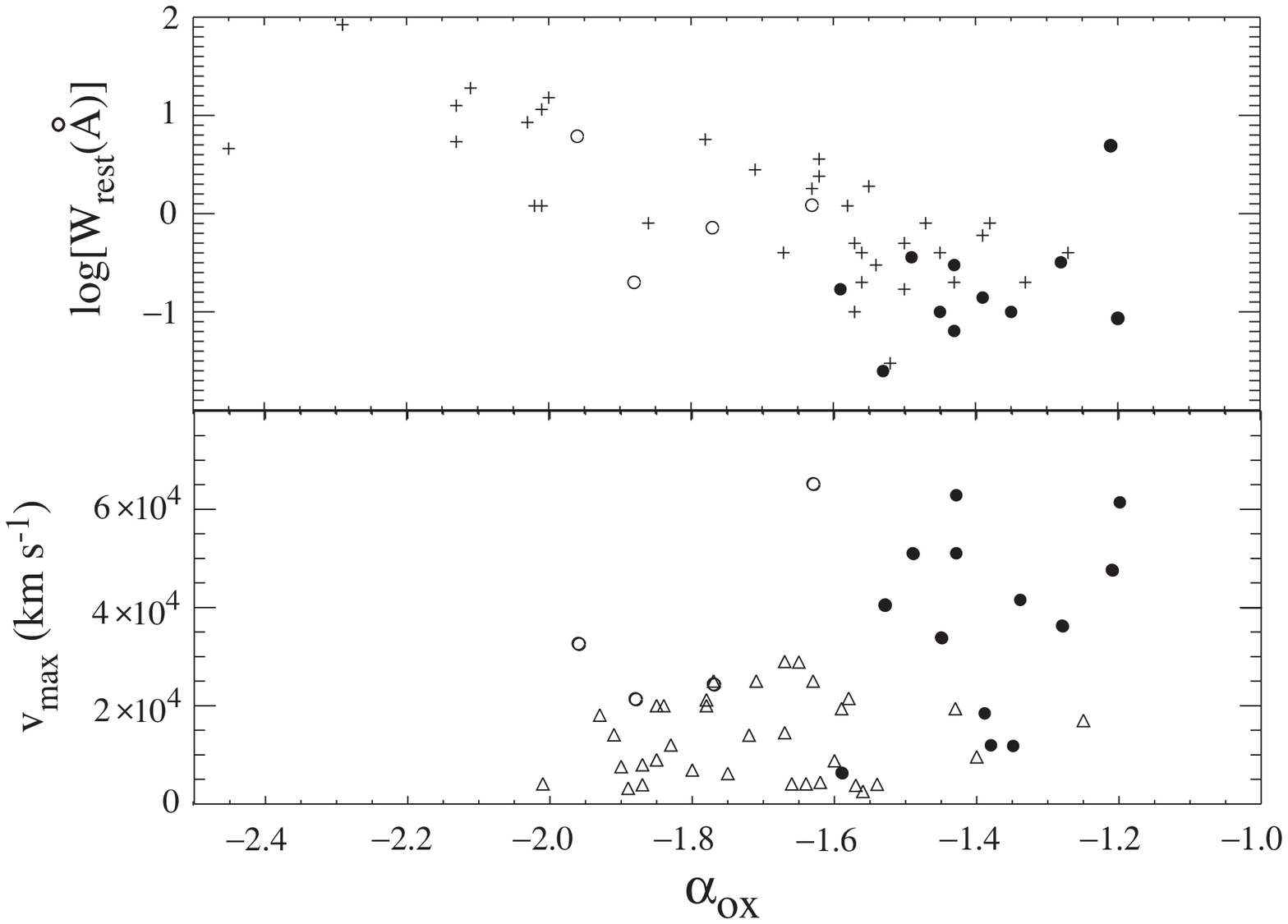}
\caption{Properties of intrinsic CIV NALs of quasars in our sample (filled circles) and the 
\cite{misawa08} sample (open circles) plotted against $\alpha_{\rm OX}$ (evaluated without corrections for intrinsic absorption). 
Top Panel: Variation of rest frame equivalent width with $\alpha_{\rm OX}$ . 
Their rest frame equivalent width is the sum of equivalent widths of all intrinsic NALs in the same 
quasar. The crosses represent the associated CIV NALs measured in low-redshift quasars 
by \cite{brandt00}. Bottom Panel: Variation of the maximum NAL velocity with $\alpha_{\rm OX}$.
The triangles are from the \cite{giustini08} BAL quasar sample.
}
\label{cartoon}
\end{figure}

\section{CONCLUSIONS}

Our results are summarized as follows:

(a) The intrinsic column densities of the X-ray absorbers in our sample of NAL quasars 
are constrained to be less than a few $\times$ 10$^{22}$~cm$^{-2}$. These values of 
$N_{\rm H}$ are consistent 
with an outflow scenario in which NAL quasars are viewed 
at smaller inclination angles than BAL quasars. 

(b) The distributions of $\alpha_{\rm OX}$ and $\Delta\alpha_ {\rm OX}$ of the NAL quasars of 
our sample differ from those of 
type 1 SDSS radio-quiet quasars and BAL quasars (Figures 2 and 3). The NAL quasars are not 
significantly absorbed in the X-ray band and the positive values of  $\Delta\alpha_ {\rm OX}$ suggest 
absorption in the UV band.

(c) The positive values of $\Delta\alpha_ {\rm OX}$ of the intrinsic NAL quasars can be explained in a geometric scenario where NAL quasars are viewed at low inclination angles (Figure 4). In this scenario, lines of sight towards a compact X-ray hot coronae of NAL quasars do not traverse the absorbing wind whereas lines of sight towards their UV emitting accretion disks do intercept the outflowing absorbers.

(d) We find that the maximum outflow velocities of the UV absorbers of NAL quasars are not correlated with their X-ray weakness in contrast to what has been reported for
BAL quasars \citep[i.e.,][]{gallagher06}.

We acknowledge financial support from NASA grants 
NNX09AB89G and NNX08AZ67G. ME, JC, and RG acknowledge support from NSF grant AST-0807993.



\end{document}